# EMF-REST: Generation of RESTful APIs from Models

Hamza Ed-douibi, Javier Luis Cánovas Izquierdo, Abel Gómez,
Massimo Tisi, Jordi Cabot

AtlanMod team (Inria, Mines Nantes, LINA), Nantes, France
`{hamza.ed-douibi,javier.canovas,abel.gomez-llana,`
`massimo.tisi,jordi.cabot}@inria.fr`

**Abstract.** In the last years, RESTful Web services have become more and more popular as a lightweight solution to connect remote systems in distributed and Cloud-based architectures. However, being an architectural style rather than a specification or standard, the proper design of RESTful Web services is not trivial since developers have to deal with a plethora of recommendations and best practices.

Model-Driven Engineering (MDE) emphasizes the use of models and model transformations to raise the level of abstraction and semi-automate the development of software. In this paper we present an approach that leverages on MDE techniques to generate RESTful services. The approach, called EMF-REST, takes EMF data models as input and generates Web APIs following the REST principles and relying on well-known libraries and standards, thus facilitating its comprehension and maintainability. Additionally, EMF-REST integrates model and Web-specific features to provide model validation and security capabilities, respectively, to the generated API. For Web developers, our approach brings more agility to the Web development process by providing ready-to-run-and-test Web APIs out of data models. Also, our approach provides MDE practitioners the basis to develop Cloud-based modeling solutions as well as enhanced collaborative support.

## 1 Introduction

Web services have increasingly become popular mainly because they simplify client/server decoupling and foster interoperability. In the last years, the emergence of distributed architectures, specially Cloud-based ones, and mobile devices have also promoted the development of lightweight applications over portable devices, which rely on web Services rather than on heavyweight desktop-based solutions.

Among the different standards for designing distributed services (e.g., SOAP, WSDL or WS-* specifications), there is a rising trend to use lightweight solutions based on plain HTTP referred to as REpresentational State Transfer (REST) [11] services. REST proposes the development of stateless distributed services and relies on simple URIs and HTTP verbs to make the Web services broadly available for a number of front-end devices. However, REST is rather an architectural style than a standard, therefore offering a considerable design and implementation flexibility. This is in theory a good thing but it can quickly lead to bad designs and architectures that end up not being conformant with the REST principles [19]. Thus, developing high-quality REST APIs for non-trivial applications may become a hard and time-consuming task.

On the other hand, Model-Driven Engineering (MDE) is a paradigm which emphasizes the use of models to raise the level of abstraction and automation in the development of software. By working at a high level of abstraction, we believe that the application of MDE to the Web environment offers important benefits in development, consistency and definition of concepts. For instance, MDE methodologies have been applied to bring more agility to the development of distributed applications by addressing the different aspects of Web development.

As a way to combine the benefits of both domains, in this paper we propose EMF-REST, an approach that leverages on MDE techniques to generate RESTful Web APIs out of plain data models. The generated RESTful Web API relies on well-known libraries and standards with the aim of facilitating its understanding and maintainability. Unlike other existing MDE-based approaches targeting the generation of Web services, ours provides a direct mapping to access data models by means of Web services following the REST principles, thus liberating Web developers from parameterizing the generation process or explicitly modeling the API to be generated. Additionally, EMF-REST takes advantage of model and Web-specific features such as model validation and security, respectively, thus providing the corresponding support in the resulting RESTful Web API. We aim to target two main communities with EMF-REST. For Web engineers, EMF-REST brings more agility to the Web development process by providing ready-to-run-and-test Web APIs out of data models. For MDE practitioners, our approach provides the basis to develop model-based solutions relying on the Cloud as well as an enhanced collaborative support for Web-based modeling tools. In this paper we will mainly focus on the first group of users.

The remainder of this paper is structured as follows. Section 2 presents some background of REST and MDE. Section 3 describes how we devised the mapping between EMF and REST principles, while Section 4 describes the additional EMF-REST features. Section 5 presents the technical architecture of the generated REST API. Section 6 describes the steps we followed to generate the API. Section 7 presents the possible applications of our approach. Section 8 discusses some related work. Finally, Section 9 concludes the paper and presents the future work.

## 2 Background

### 2.1 REST principles

In 2000, Roy Fielding identified specific design principles that led to the architectural style known as *REpresentational State Transfer* (REST) [11]. By relying on the HTTP protocol, this architectural style consists of several constraints to address separation of concerns, visibility, reliability, scalability and performance. REST principles are defined as:

**Addressable resources**  Each resource must be addressable via a Uniform Resource Identifier (URI).
**Representation-oriented**  A resource referred by one URI may have different representation formats (e.g., JSON, XML, etc.).

**Statelessness** Servers cannot hold the state of a client session. Instead, data representation formats provide information on how to manage the state of the application for each client (e.g., using embedded URIs).

**Uniform and Constrained Interface** A small set of well-defined methods are used to manipulate resources (i.e., HTTP verbs).

The last two principles are maybe the most distinguishing features of REST from other Web services specifications. According to these principles, each request is treated as an independent transaction and must only rely on the set of operations of the HTTP protocol. HTTP methods are used in REST as follows:

**GET** Used to retrieve a representation of a resource. It is a read-only, *idempotent* and *safe* operation.

**PUT** Used to update a reference to a resource on the server and it is *idempotent* as well.

**POST** Used to create a resource on the server based on the data included in the body request. It is the only *nonidempotent* and *unsafe* operation of HTTP.

**DELETE** Used to remove a resource on the server. It is *idempotent* as well.

**HEAD** Similar to GET but returning only a response code and the header associated with the request.

**OPTIONS** Used to request information about the communication options of the addressed resource (e.g., security capabilities such as CORS).

Being a collection of principles rather than a set of standards, several resources on best practices and recommendations were written to help developers to write RESTful Web services. In order to generate a high-quality RESTful Web API, we apply in EMF-REST the best practices described in [19].

### 2.2 The MDE paradigm

The MDE paradigm emphasizes the use of models to raise the level of abstraction and to automate the development of software. Abstraction is a primary technique to cope with complexity, whereas automation is the most effective method for boosting productivity and quality [25].

Modeling languages express models at different abstraction levels, and are defined by applying metamodeling techniques [7]. In a nutshell, models are defined according to the semantics of a model for specifying models, also called a *metamodel*. A model that respects the semantics defined by a metamodel is said to *conform to/to be an instance of* such a metamodel.

The Eclipse Modeling Framework (EMF) [2] has become the main reference for modeling in Eclipse [10]. Among its features, EMF allows creating metamodels – by using the Ecore language, a subset of the UML class diagrams – and their instances. Along this paper, we refer to metamodels as *Ecore models*, and their instances as *model instances*. Ecore can be considered as an implementation of MOF [20], a modeling and metamodeling standard developed by the Object Management Group (OMG). Additionally, EMF provides a generative solution which constructs Java APIs out of those models to facilitate their management, thus promoting the development of domain-specific applications.

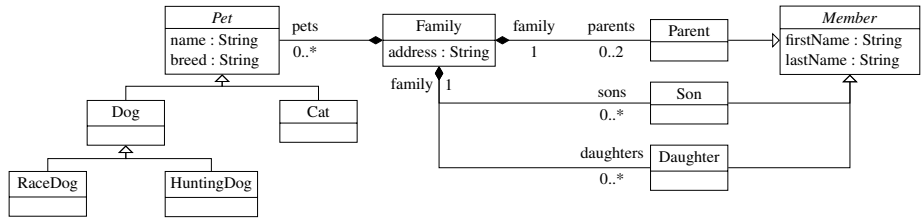

**Fig. 1.** Simple *Family* Ecore Model.

On the other hand, model transformations generate software artifacts from models, either directly by model-to-text transformations (e.g., using languages such as Acceleo or Xpand) or indirectly by intermediate model-to-model transformations (e.g., using languages such as ATL or ETL)[1]. By means of modeling languages and model transformations, it is possible to increase both the level of abstraction and provide automation in MDE, respectively.

## 3 Mapping EMF and REST principles

The first step to build EMF-REST is to align the principles behind the MDE/EMF and REST worlds. EMF-REST relies on EMF to represent the models from which the RESTful Web APIs are generated. As EMF models and their instances are managed by the corresponding APIs provided by the framework (i.e., Ecore and EObject APIs, respectively), we need to define a mapping between such APIs and the REST principles presented before. In this section we explain how we map EMF with each REST principle.

To illustrate the alignment, we will use a running example consiting in the creation of a distributed application aimed at managing information about families (e.g., parents, children, pets, etc.). Figure 1 represents a possible Ecore model for this example. As can be seen, the *Family* concept includes one attribute (i.e., *address*) to represent the address of the family and references the members (i.e., *Member* hierarchy, including *Parents*, *Sons* and *Daughters*) and pets (i.e., *Pet* concept) of the family. Additionally, different types of pets are allowed (i.e., *Cat* and *Dog* concepts plus *RaceDog* and *HuntingDog* for dog types). Figure 2 shows an intantiation of this family model for the specific case of the Simpsons family.

In what follows we will see how EMF-REST would allow to create the Simpons family by calling a REST API generated from the Family model following the REST principles.

### 3.1 Addressable Resources

Models in EMF are addressed via a URI, which is a string with a well-defined structure as shown in the expression (1). This expression contains three parts specifying: (1) a *scheme*, (2) a *scheme-specific part* and (3) an optional *fragment*. The scheme is the first part separated by the *":"* character and identifies the protocol used to access the model (e.g., *platform*, *file* or *jar*). In Eclipse we use *platform* for URIs that identify resources in Eclipse-specific locations, such as the workspace. The scheme-specific part is in the middle and its format depends on the scheme. It usually includes an *authority* that specifies a host, the *device* and the *segments*, where the latter two constitute a local path to a resource location. The optional fragment is separated from the rest of the URI by the *#* character and identifies a subset of the contents of the resource specified by URI, as we will illustrate below. The expression (2) shows an example of a platform-specific URI which refers to the Simpsons family model, represented as a file *Simpsons.xmi* contained in a project called *project* in Eclipse workspace. It is important to note that in EMF model instances include a reference to the Ecore model they conform to.

---

[1] Acceleo, XPand, ATL and ETL can be found at http://eclipse.org/modeling/

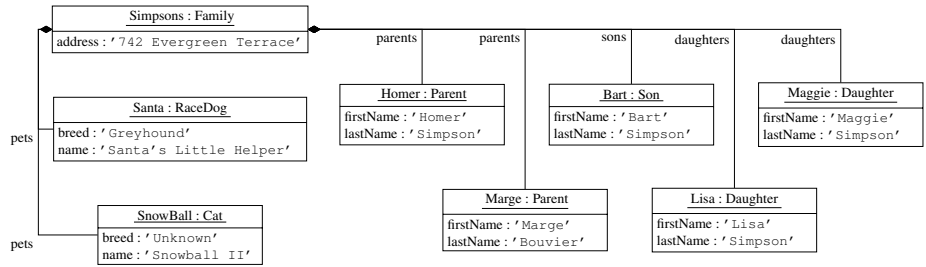

**Fig. 2.** The Simpsons Family.

$$\texttt{[scheme:][scheme-specific-part][\#fragment]} \qquad (1)$$

$$\texttt{platform:/resource/project/Simpsons.xmi} \qquad (2)$$

EMF-REST maps the previous URI to a Web URL as follows. The base URL pattern of a model instance is defined by the expression (3). In the pattern, the part *https:// [applicationLink]/rest* is the URL of the Web application, *modelId* is the identifier of the model (e.i., the Ecore model) and *ModelInstanceId* is the identifier of the model instance being accessed (the XMI file). The URL (4) represents an example to retrieve the *Simpsons* family. As can be seen, while the URI can address a file representing a model instance (where a reference to the Ecore model is included), the URL requires indicating the identifier of both the Ecore model and the model instance.

$$\texttt{https://[applicationLink]/rest/[ModelId]/[ModelInstanceId]} \qquad (3)$$

$$\texttt{https://example.com/rest/Family/Simpsons} \qquad (4)$$

This URL acts as the entrypoint for a particular model instance and points to its root element, which is normally the case in EMF. When the model instance has more that one root, EMF-REST points at the first.

Once pointing to the root of a model instance, addressing a particular element of the model in the EMF API is done by using the part *fragment* in (1). The navigation is done using the reference names in the Ecore model. For instance, the concept *Family* has the reference *parents* to access to the list of parents. Using the EMF API, the URI is shown in (5), while using the Web API, the URL is shown in (6).

$$\texttt{platform:/resource/project/Simpsons.xmi\#//@parents} \qquad (5)$$

$$\texttt{https://example.com/rest/Family/Simpsons/parents} \qquad (6)$$

Depending on the cardinality of the reference this will return a specific element – if it single-valued – or a collection of elements – if it is multi-valued (like in the case of *parents*). Accessing a specific element contained in a collection can be done using (i) the identifier of the element or (ii) its index in the list. For instance, the URI (7) retrieves the element representing *Homer* in EMF, while in EMF-REST it is done using the call (8). To identify an element, EMF-REST relies on the *identifier* flag provided by Ecore, which allows setting the attribute acting as identifier for a given class[2].

$$\texttt{platform:/resource/project/Simpsons.xmi\#Homer} \qquad (7)$$

$$\texttt{https://example.com/rest/Family/Simpsons/parents/Homer} \qquad (8)$$

On the other hand, the call (9) will retrieve the first element of the collection of parents in the EMF API. In EMF-REST, it is done by adding the parameter *index* in the URL as illustrated in the call (10).

$$\texttt{platform:/resource/project/Simpsons.xmi\#//@parents.0} \qquad (9)$$

$$\texttt{https://example.com/rest/Family/Simpsons/parents?index=0} \qquad (10)$$

### 3.2 Representation-Oriented

By default, EMF persists models using the XMI representation format. EMF-REST offers the same XMI option but also a JSON-based storage in order to comply with the representation-oriented principle of the REST architecture.

For the JSON, EMF-REST adheres to the following structure. Model concepts are represented as JSON objects containing key/value pairs for the model attributes/references. Keys are the name of the attribute/reference of the concept and values are their textual representation in one of the datatypes supported in JSON (i.e., string, boolean, numeric, or array). For attributes, their values are mapped according to the corresponding JSON supported datatype or String when there is not a direct correspondence (e.g., float-typed attributes). When the attribute is multi-valued, its values are represented using the array datatype. For references, the value is the URI of the addressed resource within the server (if the reference is multi-valued, the value will be represented as an array of URIs). Listing 1 shows an example of the content format in JSON. Note that references containing a set of elements from model hierarchies are serialized as a list of JSON objects corresponding to their dynamic type (see *pets* reference including *Race-Dog* and *Cat* JSON objects).

In XML, model concepts are represented as XML elements including an XML element for each model attribute/reference. Attribute values are included as string values

---

[2]When the *identifier* flag is not used, the fallback behavior looks for an attribute called *id*, *name* or having the *unique* flag activated.

**Listing 1.** Partial JSON representation of the Simpsons family.

```
 1  {
 2    "family":{
 3      "address":"742 Evergreen Terrace",
 4      "parents":{
 5        "parent":[{
 6          "uri":"https://example.com/rest/Family/Simpsons/parents/Homer"},{
 7          "uri":"https://example.com/rest/Family/Simpsons/parents/Marge"}]
 8      },
 9      "pets":{
10        "raceDog":{"uri":"https://example.com/rest/Family/Simpsons/pets/Santa's
               Little Helper"},
11        "cat":{"uri":"https://example.com/rest/Family/Simpsons/pets/Snowball II"}
12      }
13      ...
14    }
15  }
```

**Listing 2.** Partial XML representation of the the Simpsons family.

```
 1  <family>
 2    <address>742 Evergreen Terrace</address>
 3    <parents>
 4      <parent>
 5        <uri>https://example.com/rest/Family/Simpsons/parents/Homer</uri>
 6      </parent>
 7      <parent>
 8        <uri>https://example.com/rest/Family/Simpsons/parents/Marge</uri>
 9      </parent>
10    </parents>
11    <pets>
12      <raceDog>
13        <uri>https://example.com/rest/Family/Simpsons/pets/Santa's Little Helper
               </uri>
14      </raceDog>
15      <cat>
16        <uri>https://example.com/rest/Family/Simpsons/pets/Snowball II</uri>
17      </cat>
18    </pets>
19    ...
20  </family>
```

in the XML element representing such attribute, references are represented according to their cardinality. If the reference is single-valued, the resulting XML element will include only the URI of the addressed resource in the server. On the other hand, if the reference is multi-valued, the resulting XML element will include a set of XML elements including the URIs addressing the resources. Listing 2 shows an example of the content format in XML format.

### 3.3 Uniform and Constrained Interface & Statelessness

EMF supports loading, unloading and saving model instances after their manipulation. In EMF-REST, these operations are managed by the application server. Models are loaded (and unloaded) dynamically as resources when running the application managing the Web API, and they are saved after each operation is done, thus conforming to the REST statelessness behavior.

**Listing 3.** Update the attribute of a concept using EMF generated API.

```
1  ...
2  hommerObj.setName("Homero"); // hommerObj is of type Parent
3  ...
```

**Listing 4.** HTTP call and JSON representation to update the firstname of the addressed parent.

```
1  PUT https://example.com/rest/Family/Simpsons/parents/Homer
2
3  {"parent":{
4          firstName:"Homero"
5          }
6  }
```

**Table 1.** Supported operations in the generated API.

| Operation | HTTP Method | URL | Model |
|---|---|---|---|
| CREATE and add element to the collection | POST | `.../a/bs` | |
| READ all the elements from the collection | GET | | |
| READ the element (1) identified by <id>, (2) in the <i> position of the collection, or (3) the element c | GET | | |
| UPDATE the element (1) identified by <id>, (2) in the <i> position of the collection, or (3) the element c | PUT | (1) `.../a/bs/<id>` (2) `.../a/bs?index=<i>` (3) `.../a/c` | |
| DELETE the element (1) identified by <id>, (2) in the <i> position of the collection, or (3) the element c | DELETE | | |

To manipulate model instances, EMF enables the basic CRUD (i.e., create, read, update and delete) operations over model instances by means of either the EMF generated API or the EObject API. In EMF-REST, the same CRUD operations are mapped into the corresponding HTTP methods (*POST*, *GET*, *PUT*, and *DELETE*). For instance, Listing 3 shows the code to modify the name of the parent called *Homer* using EMF generated API for the Family model. The same operation can be done on our EMF-REST API by sending the PUT HTTP method containing the JSON representation of the new *Parent* model element, as shown in Listing 4.

Table 1 shows how each CRUD operation is addressed along with several URL examples. The first column of the table describes the operations. As can be seen, the first two rows represent operations over collections, enabling adding new elements (see first row) and reading their content (see second row). The rest of the rows describe operations over either individual elements of a collection (see cases 1 and 2 of these operations) or elements contained in a single-valued reference (see case 3). The second column shows the correspondent HTTP method for each operation while the third column presents the corresponding URL for each case. Finally, the last column includes a small model to better illustrate the cases considered in the table.

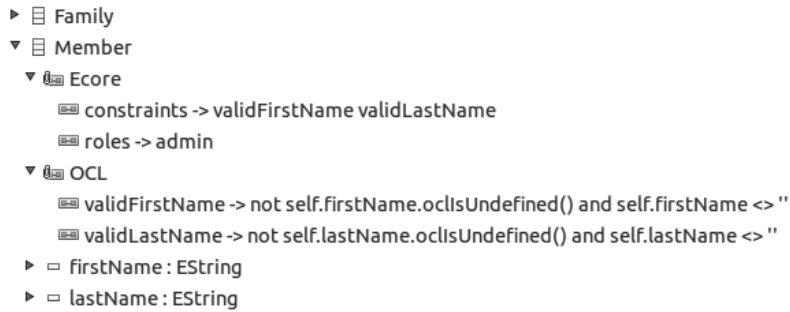

**Fig. 3.** Annotations on an excerpt of the example model.

## 4 Additional EMF-REST Features

EMF-REST also provides support for validation and security aspects in the generated RESTful Web API. With this aim, our approach leverages on model and Web-specific mechanisms, as we will describe below.

### 4.1 Validation

Validation is the process of verifying that the given data inputs respect the defined constraints at the model level Support for validating the API data calls is pretty limited in current web technologies. The most relevant one for our scenario would be the Bean Validation specification that defines an API to enforce the validation of *Java Beans*. However, this specification can only ensure that fields follow certain constraints (e.g., a field is not `null`) and cannot satisfy complex validation scenarios for model integrity (e.g., a *son* cannot have more than two parents). On the other hand, MDE provides specific support for validating models, for instance, by providing an implementation (both IDE support and run-time engine) the Object Constraint Language [26], a language complementing UML [23] that allows software developers to write complex constraints over object models. Thus, for validation, EMF-REST employs a model-based approach using OCL to define constraints as annotations in the model elements.

OCL annotations can be attached to concepts in the model as invariants. An example on the Simpsons family is shown in Figure 3. As can be seen, concepts include a set of invariants inside the annotation *OCL* plus the annotation *Ecore/constrains* which specifies the invariants to execute. Invariants are checked each time a resource is modified (i.e., each time the Web API is called from a Web-based client using the `POST`, `PUT` or `DELETE` methods). This validation scheme is imposed to comply with the stateless property of REST architetures, however, it may involve some design constraints when creating the model. In those cases where models cannot be validated each time they are modified (e.g., creating model elements requires several steps to fulfill cardinality constraints), EMF-REST allows this validation process to be temporary be deactivated. The results of the validation process are mapped into the corresponding HTTP response messages (i.e., using status codes)

## 4.2 Security

Security is one of the cornerstones when developing Web-enabled applications. In this case, there is little support for security definition and enforcement from the MDE side since most tools just execute on a local environment but we have plenty of support from web technologies to add security aspects to our generated API and protect the access to the data models. In particular, our approach allows designers to provide some security annotations on the model that are then translated into security restrictions as described below. As part of the generation, EMF-REST also creates a separated admin view where additional security information (like users and passwords) can be maintained.

In order to secure a Web application, we have to: (i) ensure that only authenticated users can access resources, (ii) ensure the confidentiality and integrity of data exchanged by the client and the server from the moment of sending a request to the moment of receiving a response, and (iii) prevent unauthorized clients from abusing data. In order to address the previous requirements, EMF-REST relies on a set of security protocols and services provided by *Java EE* which enable encryption, authentication and authorization in Web APIs, as we will explain in the following.

**Encryption** If the connection is not secured, it is possible to intercept the packets and collect sensitive data when interacting with a Web API. Encrypting the information exchanged between the client and server makes it only readable by the destination holding the decryption key. The Web defines HTTPS protocol to add the encryption capacities of SSL/TLS to standard HTTP communication. EMF-REST enforces the use to HTTPS to communicate with its services.

**Authentication** Authentication is about validating the identity of a client attempting to access a resource on a server. The validation process checks if the client has provided valid credentials and can be performed using several protocols, namely: basic authentication, digest authentication and X.509 certificates. EMF-REST relies on basic authentication to provide the authentication mechanism since it is simple, widely supported, and secure by using HTTPS. The basic authentication involves sending a Base64-encoded username and password within the HTTPS request header to the server. Upon reception, the server checks the credentials and send a response containing the requested data if credentials are correct or an unauthorized response otherwise.

**Authorization** The role of authorization is to define if the user has the permission to access a resource. Authorization is usually combined with authentication to identify users and determine which type operation is allowed for each one of them. While the authentication is enabled by the protocol/server, the authorization is generally provided by the application, which knows the permissions for each operation in the resources. EMF-REST uses a simple role-based mechanism to support authorization in the generated Web API. Roles are associated to users (i.e., authentication) and operations in the Web API (i.e., authorization). In EMF-REST roles are assigned to resources by adding annotations to the model. Figure 3 illustrates the use of these annotations (e.g., see annotation *Ecore/roles* in the *Member* concept). Additionally, EMF-REST also provides an admin view to create users and assigns one or more roles to each user, as we will show in section 5.3.

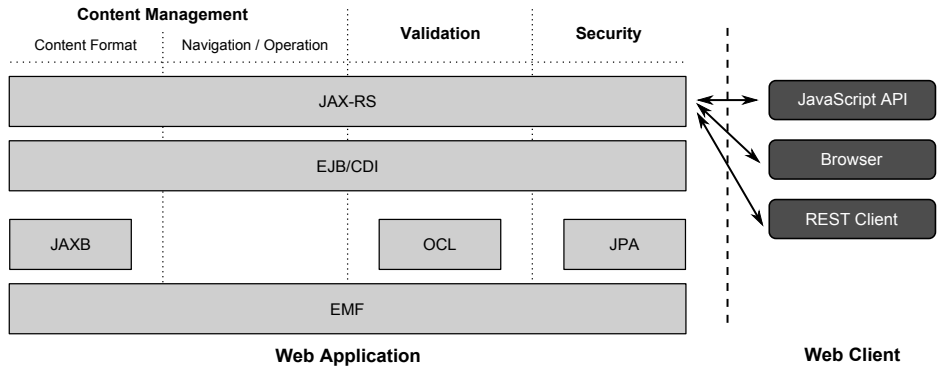

**Fig. 4.** Architecture of EMF-REST generated application.

## 5 EMF-REST API Architecture

To implement the features described in the previous sections, we devised the application architecture presented in Figure 4. This architecture can then be seamlessly accessed with a variety of clients.

The Web application is split into three main components according to the functionality they provide: (1) content management, (2) validation and (3) security. The application relies on EMF as modeling framework and uses the following additional frameworks/specifications for each component, respectively: (1) *Java Architecture for XML Binding (JAXB)* to enable the content format support, (2) Eclipse OCL framework to provide validation before updating the model, (3) *Java Persistence API (JPA)* to provide security support by storing the system users and their permissions in an embedded database. The Web application also leverages on *Enterprise Java Bean (EJB)*, *Context dependency Injection (CDI)* and *Java API for Representational State Transfer (JAX-RS)* specifications. EJBs enable rapid and simplified development of distributed, transactional, secure and portable applications. They are in charge of loading the EMF resources from the persistent storage and providing the necessary methods to manage the resources (e.g., obtaining objects from the resource, removing objects) in a secure and transactional way. These EJBs are then injected into JAX-RS services using CDI technology. Thus, JAX-RS is used to expose EMF resources as Web services. In the remaining of the section we describe how all these technologies are used in each component.

### 5.1 Content Management

This component addresses the mapping between EMF and REST principles. It is in turn split into two subcomponents: (1) content format, which addresses the mapping of the second REST principle (i.e., Representation-Oriented); and (2) navigation/operation, which addresses the rest of the REST principles.

Regarding the content format, EMF-REST enriches the EMF-generated API with JAXB[3] annotations, which enable the support for mapping Java classes to XML/J-

---

[3] https://jaxb.java.net/

**Listing 5.** Part of the Family concept.

```java
1   @XmlRootElement (name="family")
2   public class FamilyImpl extends EObjectImpl implements Family {
3
4       protected EString address;
5       protected EList<Parent> parents;
6
7       @XmlElementWrapper(name = "parents")
8       @XmlJavaTypeAdapter(ParentAdapter.class)
9       @XmlAnyElement(lax=true)
10      public EList<Parent> getParents() {
11          if (parents == null) {
12              parents = new EObjectContainmentWithInverseEList<Parent>(Parent.
                    class, this, ExamplePackage.FAMILY__PARENTS, ExamplePackage.
                    PARENT__FAMILY);
13          }
14          return parents;
15      }
16
17      @XmlElementWrapper(name = "pets")
18      @XmlJavaTypeAdapter(PetAdapter.class)
19      @XmlAnyElement(lax=true)
20      public EList<Pet> getPets() {
21          if (pets == null) {
22              pets = new EObjectContainmentEList<Pet>(Pet.class, this,
23                  ExamplePackage.FAMILY__PETS);
24          }
25          return pets;
26      }
27
28      @XmlElement
29      public String getAddress()  {
30          return address;
31      }
32      ...
33
34  }
```

SON (i.e., marshalling/unmarshalling Java object into/from XML/JSON documents). The Listing 5 shows an example of the use of JAXB annotations to produce the corresponding representation in JSON (as shown in Listing 1) and XML ((as shown in Listing 2). As can be seen, each concept class is mapped to an `XmlRootElement` element, while either `XmlElement` or `XmlElementWrapper` elements are used to map the attributes or references of the class, respectively. Other annotations are used to deal with the references and inheritance. For instance, `XmlJavaTypeAdapter` is used to associate a reference of an element with the correspondent representation.

Navigation and operations are enabled by using JAX-RS, which provides a set of Java APIs for building Web services conforming to the REST style. Thus, this specification defines how to expose POJOs as Web resources, using HTTP as network protocol. For each concept (e.g., *Family*) a resource will be created (e.g., `FamilyResource`) annotated with `Path` (e.g., `@Path("Family")`). The `@Path` annotation has the value that represents the relative root URI of the addressed resource. For instance, if the base URI of the server is `http://example.com/rest/`, the resource will be available under the location `http://example.com/rest/Family`. To produce a particular response when a request with GET, PUT, POST and DELETE is inter-cepted by a resource, resource methods are annotated with `@GET`, `@PUT`, `@POST` and `@DELETE` what are invoked for each corresponding HTTP verb.

### 5.2 Validation

EMF-REST leverages on Eclipse OCL[4], which provides an implementation of the OCL OMG standard for EMF-based models, to validate the data by means of annotations including the constrains to check the model elements. The generated API relies on the provided APIs for parsing and evaluating OCL constraints and queries on Ecore models. When constraints are not satisfied, the validation process will fire an exception that will be mapped by JAX-RS into an HTTP response including the corresponding message indicating the violated constraint.

### 5.3 Security

EMF-REST relies on the combination of Java EE and JAX-RS for the authentication and authorization mechanisms by using the concept of role, while encryption is provided via HTTPS. To enable authentication, the deployment descriptor of the WAR file (i.e., `WEB-INF/web.xml`) has been modified to include the security constraints (i.e., `<security-constraint>`) defining the access privileges. Assigning permissions for HTTP operations based on the roles provided in the model is done by using the `@RolesAllowed` annotation. For example, as shown before, Figure 3 shows that the role allowed for the *Member* concept is admin. This will restrict access to the resource to the users having the role `ADMIN`. To express this in the generated API, the annotation `@RolesAllowed({"ADMIN"})` is placed on top of `MemberResource`. If no role is assigned to a concept, a `@PermitAll` annotation is placed on the resource class meaning that all security roles are permitted to access this resource. Note that security roles assigned to a resource are not inherited by its sub-resources.

To manage the list of users and their roles, EMF-REST generates an admin view that allows the manager of the API to add, edit and remove users. All created users have a default role (i.e., *user*) allowing them to access unannotated concepts. The manager can assign more roles to a user in order to grant him/her access to a specific resource.

## 6 Code Generation and Tool Support

In order to generate the REST APIs we created a Java tool available as an open-source Eclipse plugin [1]. Figure 5 shows the steps followed by the tool to generate the application starting from an initial Ecore model.

Step I of the process generates a Maven-based [5] project that serves as a skeleton of the application. Maven allows a project to be built by using the so called *Project Object Model* (POM) file, thus providing a uniform build system. The POM is initialized with the required library dependencies described in the previous section.

---

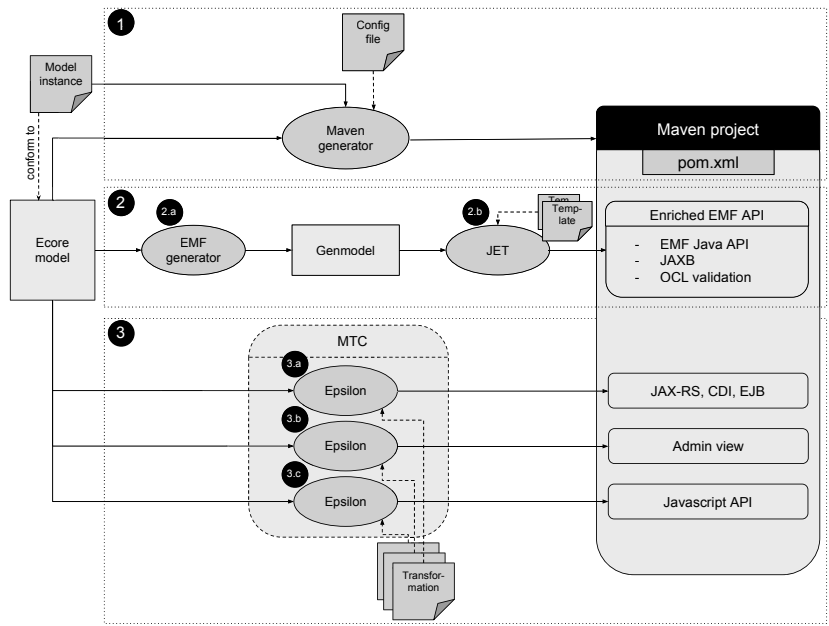

**Fig. 5.** EMF-REST generation process.

In Step II, the EMF code generation facility has been extended to include the required support for JAXB and validation. In particular, the EMF generator templates have been extended to produce the code corresponding to the JAXB annotations and the required methods to execute the OCL validation process.

Step III performs a set of model-to-text transformations using the ETL[6] language to generate the remaining elements, including: (1) the JAX-RS, CDI and EJB implementation classes, (2) the admin view developed and (3) a simple JavaScript API to facilitate Web developers to build clients for the generated Web API. For each part of the application (e.g., JAX-RS resources, etc.), an ETL transformation template has been implemented to generate the appropriate behavior according the input Ecore class. Since this step requires several transformations, the MTC tool [5] has been used to orchestrate the flow of the *Epsilon* templates.

# 7 Applications

In this section we discuss potential applications of EMF-REST. The first section covers the main scenario targeting EMF-REST illustrating how our approach can help web developers to quickly create/prototype web services, while the second section illustrates how EMF-REST can set the basis for promoting scalability and collaboration in MDE.

---

[6]https://www.eclipse.org/epsilon/

### 7.1 Generation of Web Services

To develop a REST API for a data model, web developers have to implement a middleware framework supporting the REST access to the model elements and, additionally, develop the extra support to cover a number of typically required features such as security, persistence or validation. Thus, the development process requires a considerable effort, not only to implement the services, but also to study and apply the different specifications and standards. When applying EMF-REST, designing Web services boils down to providing the required data model as input and running the code generation suite. Thus, EMF-REST can be used to obtain the core RESTful Web API allowing developers to focus more on designing the REST client at an early stage of the development process of a system. Even for web developers preferring to develop the REST API themselves at a later stage, our generated one would be useful for prototyping and validation purposes.

### 7.2 Scalability and Collaboration in MDE

Beside the application in the Web realm, EMF-REST also provides the basis for Web-based model management. Many benefits can be achieved by moving model management from classic environments (i.e., the Eclipse IDE) into distributed (i.e., Cloud) architectures by the use of Web services. For instance, to cope with the increase of MDE applications to large and more complex systems, which has led to performance issues [16]. By using EMF-REST, developers have the ground to deploy the system in a Platform-as-a-Service (PaaS) provider, in which model management would take advantage of its scalability capabilities and may solve performance issues in MDE systems. Thus, developers would not be limited by the capacity of their local machines in term of memory and CPU. For instance, resource and time-consuming operations such as validation or transformation would benefit from the scalable infrastructure of the current Cloud providers (e.g., Amazon WS, Openshift).

Adopting a distributed system would also promote the collaboration between modelers. As modeling activities rise and developer teams get geographicaly dispersed, collaborative modeling environments are more and more necessary [4]. Using EMF-REST at the metamodel level (i.e. the input of EMF-REST would not be a data model but a metamodel framing the kind of models that could be created by calling the API methods) would promote the collaborative development of new software models for the project at hand or even the development of Domain-Specific Languages (DSLs). For example, developers can work on a slice of the DSL while the others can provide live feedbacks to improve and polish the DSL under development. Approaches such as Collaboro [8], currently providing a centralized collaboration environment, could benefit from EMF-REST features to offer a distributed solution.

## 8 Related Work

Several efforts have been made to bring together MDE and Web Engineering. This field is usually referred to as Model Driven Web Engineering (MDWE) and proposes the use of models and model transformations for the specification and semiautomatic generation of Web applications  [24] [14] [21] [9] [27] [15] [13] [12] [17]. Mainly, data models, navigation models and presentations models are used for this purpose.

Some of these works provide support for the generation of Web services as well, but support for generation of RESTful APIs is very limited [22],[3] and [18]. Moreover, these approaches require the designer to specifically model the API itself using some kind of tool-specific DSL from which then the API is (partially) generated. Instead, our approach is able to generate a complete RESTful API implementation from a plain data model.

## 9   Conclusion

In this paper we have presented EMF-REST, an approach to generate ready-to-run-and-test RESTful Web APIs out of domain models. The generated APIs rely on well-known libraries and standards, and also provide extra features such as validation and security. We believe our approach fills an important gap between the Web and modeling technologies, thus enabling Web developers to leverage on modeling techniques to generate RESTful Web APIs and MDE practitioners to bring their models into the Web. EMF-REST has been released as an Eclipse-plugin and is available at [1].

As further work, we would like to work on a small configuration DSL to help designers parameterize the style of the generated API (e.g., configuring the URIs to the resources). Existing approaches, like WADL[7] and RSDL[8], which also propose DSLs to describe Web APIs, could be useful here. We are also interested in evaluating the security and scalability of our approach and devise possible improvements (e.g., a more fine-grained security mechanism or the implementation of different types of model storage, e.g., NoSQL-based ones like Neo4EMF [6], at the server side). Last but not least, we would also like to explore the benefits of using EMF-REST in combination with client-side modeling environments, for instance, in Eclipse, thus enabling developers to deal with large EMF models in a transparent way (i.e., models in Eclipse that are remotely stored using an EMF-REST backend).

---

[7] http://www.w3.org/Submission/wadl/

[8] http://goo.gl/7wpf9y